\documentclass[conference]{IEEEtran}

\usepackage{dsfont}      
\usepackage{amssymb}      

\usepackage{amsmath}
\usepackage{graphicx}
\usepackage{subfigure}
\ifCLASSINFOpdf
\else
\fi
\begin{document}
%
\title{Modeling and Predicting DNS Server Load}



\author{\IEEEauthorblockN{Zheng Wang}
\IEEEauthorblockA{Information Technology Laboratary\\
National Institute of Standards and Technology\\
Gaithersburg, MD 20899, USA\\
Email: zhengwang98@gmail.com}
}


%


\maketitle

\begin{abstract}
The DNS relies on caching to ensure high scalability and good performance. In optimizing caching, TTL adjustment provides a means of balancing between
query load and TTL-dependent performances such as data consistency, load balancing, migration time, etc.
To gain the desired balance, TTL adjustment depends on predictions of query loads under alternative TTLs.
This paper proposes a model of DNS server load, which employs the uniform aggregate caching model to simplify the complexity of modeling clients' requests and their caching.
A method of predicting DNS server load is developed using that model.
The prediction method is solely based on the unilateral measurements or observations at authoritative servers. Without reliance on
lots of multi-point measurements nor distributed measuring facilities, the method is best suited for DNS authoritative operators.
The proposed model and prediction method are validated through extensive simulations. Finally, global sensibility analysis is conducted to evaluate the impacts of measurement uncertainties or errors on the predictions.
\end{abstract}


%
\IEEEpeerreviewmaketitle

\section{Introduction}
The Domain Name System (DNS) provides an indispensable substrate for Internet applications and services by linking human-friendly names with machine-readable addresses. Being a deceptively simple protocol based on the client-server model, the DNS basically forwards users' queries to authoritative servers, asking for the translation from name to address or vice visa, which then return the answers. Given strenuous efforts of optimizing the DNS efficiency, security and privacy, the DNS is increasingly complex in its implementation and operation.

A DNS client typically does not directly interact with authoritative server. Rather, it relies on a caching resolver to iteratively query the relevant authoritative servers for the final answer. This model intentionally makes a simple and lightweight DNS client, leaving caching resolver tasked with iteration, caching, and even validating on behalf of DNS client.

In the client-resolver-authoritative model, DNS caching performed at caching resolver also enables scalability in terms of authoritative server's load. With caching, a DNS record recently retrieved by a caching resolver can be temporarily kept in cache for future responding. Hence a caching resolver may find cache hit for a large population of incoming recursive queries, rather than simply issue an outstanding query for each of them. Since original DNS queries from clients may be greatly filtered by caching resolver, the resultant query load on authoritative server and the query latency would be reduced. This ensures the DNS scales well with surging Internet terminals and intensified DNS usages.

The duration of DNS caching follows a Time-to-Live (TTL) specified as a field of DNS record. The initial value of TTL decreases with time since the record's arrival at a caching resolver. The record in cache is evicted and no longer used for immediate response once its TTL value becomes zero. Naturally, a large TTL means a high cache hit rate and thereby an attenuated query load on authoritative server. In general, the higher the TTL, the less frequently the authoritative server is accessed. However, the penalty of a large TTL is the slow propagation speed of authoritative DNS records to caching resolver. Given the TTL-based caching mechanism, a caching resolver has no opportunity to refresh a record in cache until its TTL expires. The cached copies maintained at caching resolver may be inconsistent with the authoritative records served at authoritative server. A DNS update on the authoritative side is unlikely to be immediately propagated to the caching side. In general, the higher the TTL, the slower DNS updates propagates. Therefore, any TTL value reflects a balance between server load (as well as query latency) and fresh data.

In the DNS operation practice, TTL often needs to be adjusted or reset for different purposes in variant scenarios. For example, the migration of a site may result in change of the site server's IP address, which requires an update on the associated DNS record mapping the site's domain to its IP address. Due to cache inconsistency, some users, misled by the cached old record, may still navigate to the old server after the migration and thus experience the downtime. Since the maximum downtime can be approximated by the TTL, the DNS operator may reduce the TTL for a minimum downtime prior to the migration and then restore the original TTL after the migration. Another example of lowering the TTL is the fine-grained DNS based load balancing. As probably the most common nonproxy load distribution strategy used by applications, Round-Robin DNS are employed to balance the workload of multiple application servers accessed via one domain name. The IP address of each application server is provisioned in one record of the domain name's record set. The load balancing preferences among application servers are encoded by the order of records which is controlled by authoritative server operator. The effectiveness of DNS based load balancing is distorted by the caching effects. The higher the TTL, the less fine-grained load balancing. So DNS operator may tune the TTL towards low or even zero for a better load balancing. The major concern over aggressively decreasing TTL is the effect of increasing the load on authoritative server, which may overload or overwhelm authoritative server. Conversely, when an authoritative server detects heavy request traffic which almost saturates its capability, it may choose a higher TTL as a means of defense to ensure the availability and resilience. In all of the TTL tuning scenarios discussed above, DNS operators are always confronted with the problem of predicting DNS server load under an alternative TTL.

DNS server load is the aggregate DNS request rate originated from individual client, filtered through caching, and destined to the target domain. The complexity of predicting DNS server load can be reflected in three aspects:

$\bullet$ The knowledge about each individual client's request pattern is hardly known in practice. Since each individual client is hidden from the authoritative server by the caching resolver, the authoritative server can neither count each individual client nor figure out its request rate.

$\bullet$ The profile regarding what individual client is served by what caching resolver is invisible to the authoritative server. The authoritative server can only observe the query traffic as output of each caching resolver, not the original query traffic as input of each caching resolver.

$\bullet$ The caching mechanism is hard to quantitatively model in terms of the collective effect of request filtering by each caching resolver. While there were a number of measurement and modeling studies in recent years, their focus limits to a single caching resolver, not aggregated caching resolvers.

This paper proposes an uniform aggregate caching model to simplify the analysis of equivalent aggregate query load from a diversity of nonuniform clients. Using the proposed model, the complex DNS query pattern with caching effects is parameterized. The load of authoritative servers can be obtained by retrieving the parameter(s) of the model using the limited observations from the authoritative servers.

The rest of this paper is organized as follows.

\section{The Prediction Method}

\begin{figure}[!t]
\centering
\includegraphics[width=3.2in]{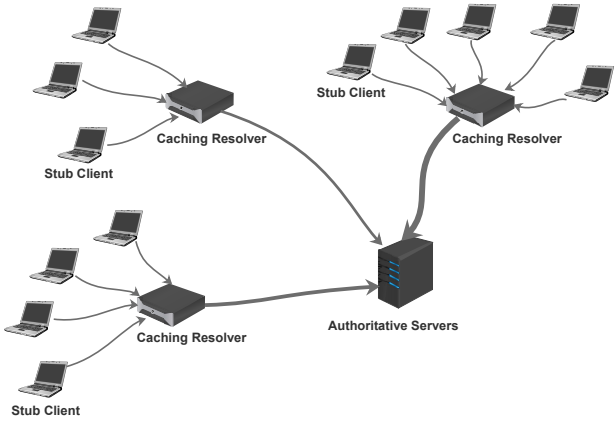}
\caption{Original stub client only DNS model.}
\label{Part-1}
\end{figure}

\begin{figure}[!t]
\centering
\includegraphics[width=3.2in]{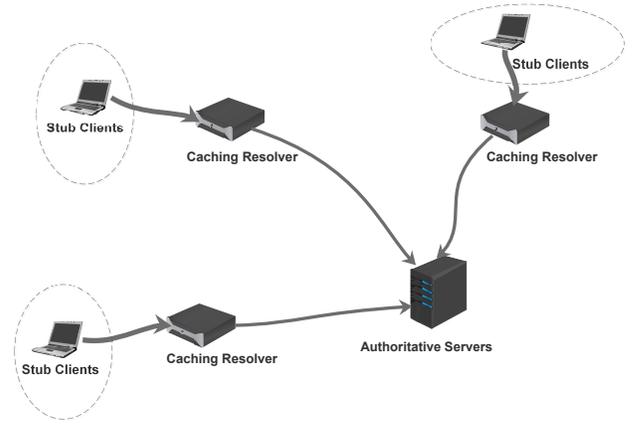}
\caption{Equivalent stub client only DNS model using the UAC model.}
\label{Part-1}
\end{figure}

\begin{figure}[!t]
\centering
\includegraphics[width=3.2in]{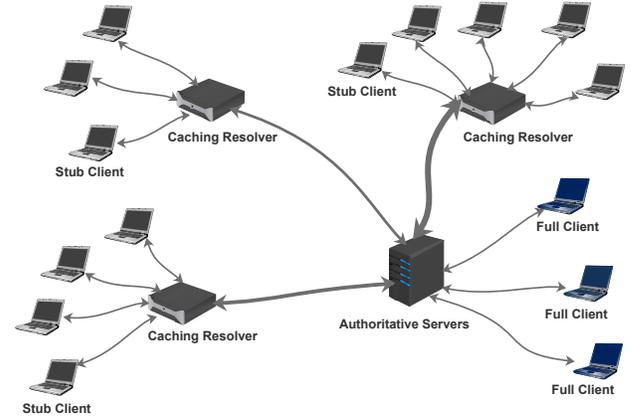}
\caption{Original stub client and full client DNS model.}
\label{All-1}
\end{figure}

\begin{figure}[!t]
\centering
\includegraphics[width=3.2in]{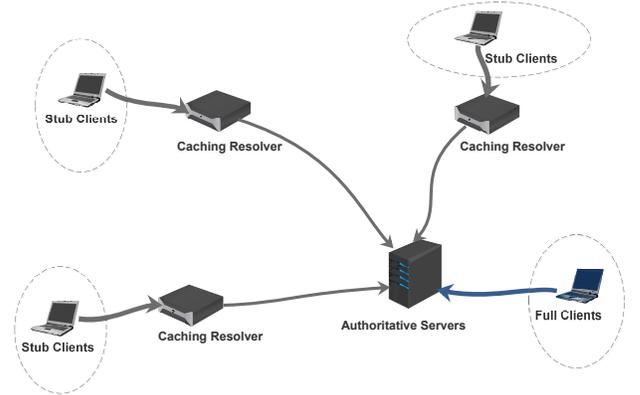}
\caption{Equivalent stub client and full client DNS model using the UAC model.}
\label{All-1}
\end{figure}

\subsection{Stub Client Only Model}

We first assume a client-resolver-authoritative model here. The TTL of the requested DNS record is set as $\tau$.

There are $N$ caching resolvers, $\mathds{R}_1, \mathds{R}_2, ..., \mathds{R}_N$ querying the authoritative servers for the DNS record. Caching resolver $\mathds{R}_i$ serves $M_i$ stub clients, $\mathds{S}^i_1, \mathds{S}^i_2, ..., \mathds{S}^i_{M_i}$, $i=1,2,...,N$. The request rate of stub client $\mathds{S}^i_j$ for the DNS record is $a^i_j$, $i=1,2,...,N$, $j=1,2,...,M_i$.

The incoming DNS request rate for the DNS record observed by caching resolver $\mathds{R}_i$, $i=1,2,...,N$ can be aggregated as

\begin{equation}
\mathcal{A}_i=\sum_{j=1}^{M_i}a^i_j
\end{equation}

And with some requests hit in cache, the outgoing DNS request rate for the DNS record from caching resolver $\mathds{R}_i$, $i=1,2,...,N$ towards the authoritative servers is $b_i$, $i=1,2,...,N$.

Note that for constant stub clients and their invariant request rates, $\mathcal{A}_i$ can be considered as constant. Therefore, the outgoing request rate of each individual caching resolver is simply determined by the TTL-based caching mechanism. That dependency can be expressed as

\begin{equation}
b_i(\tau)=\frac{\mathcal{A}_i}{1+\mathcal{A}_i*\tau}
\end{equation}

The incoming DNS request rate observed by authoritative servers is the aggregate of the outgoing DNS request rates from all caching resolvers, and thus can be written as

\begin{equation}
\mathcal{B(\tau)}=\sum_{i=1}^{N}b_i(\tau)
\end{equation}

The authoritative server is commonly able to and authorized to observe, record, and analyze the incoming DNS requests. By identifying the source IP address of each DNS message, the authoritative server can figure out every caching resolver as well as its request rate. So the server load and the count of caching resolvers are known by the authoritative server. Since the TTL is totally controlled by the authoritative server, the TTL value is also available.

Because the profile of individual clients belonging to each caching resolver is always unknown, we simplify the aggregate model as the uniform aggregate caching (UAC) model. The UAC model assumes that the original aggregate request rate as input of caching resolver equally distributes among all caching resolvers. That is

\begin{equation}
\mathcal{A}_1=\mathcal{A}_2=,...,=\mathcal{A}_N=\widetilde{\mathcal{A}}
\end{equation}

Because of Eq(2), we also have

\begin{equation}
b_1(\tau)=b_2(\tau)=,...,=b_N(\tau)=\widetilde{b}(\tau)
\end{equation}

So there will be $N$ uniformly requested caching resolvers in the UAC model as equivalent to the $N$ diversely requested caching resolvers in practice. So the query load of authoritative servers can be written as

\begin{equation}
\widetilde{\mathcal{B}}(\tau)=\frac{N*\widetilde{\mathcal{A}}}{1+\widetilde{\mathcal{A}}*\tau}
\end{equation}

Assume the constant query pattern of stub clients and the constant requesting caching resolvers.

Given one TTL of the requested record $\tau_0$ and the respective query load of authoritative servers $\widetilde{\mathcal{B}}(\tau_0)$, we can derive the equivalent aggregate request rate arriving at each caching resolver as

\begin{equation}
\widetilde{\mathcal{A}}=\frac{\widetilde{\mathcal{B}}(\tau_0)}{N-\widetilde{\mathcal{B}}(\tau_0)*\tau_0}
\end{equation}

For a new TTL of the requested record $\tau^*$, the query load of authoritative servers can be predicted using the estimation $\widetilde{\mathcal{A}}$:

\begin{equation}
\widetilde{\mathcal{B}}(\tau^*)=\frac{N*\widetilde{\mathcal{A}}}{1+\widetilde{\mathcal{A}}*\tau^*}
\end{equation}

\subsection{Stub Client and Full Client Model}

Besides the stub clients and their respective caching resolvers assumed above in the stub client only model, we consider some full clients simultaneously requesting the authoritative servers. Here full clients are able to contact the authoritative servers by themselves, independent of caching resolvers. Hence the authoritative servers will observe not only the incoming query traffic from caching resolvers but also that from full clients. Note that the former is unilaterally impacted by the TTL of the requested DNS record, whereas the latter is obviously not related to that TTL at all. Based on the assumptions in the stub client only model, we further add $K$ full clients, namely $\mathds{C}_1, \mathds{C}_2, ..., \mathds{C}_K$. The request rate of full client $\mathds{C}_i$ for the DNS record is $c_i$, $i=1,2,...,K$.

The incoming DNS request rate observed by authoritative servers is the aggregate of the outgoing DNS request rates from all caching resolvers plus the overall DNS request rates from all full clients:

\begin{equation}
\mathcal{B}^{'}(\tau)=\sum_{i=1}^{N}b_i(\tau)+\mathcal{D}
\end{equation}

Where $\mathcal{D}$ is the overall query rate from full clients:

\begin{equation}
\mathcal{D}=\sum_{i=1}^{K}c_i
\end{equation}

Similar to the stub client only model, we also suppose $N$ equivalent uniformly requested caching resolvers so that Eq (4) and (5) holds. The query load of authoritative servers can be expressed as

\begin{equation}
\widetilde{\mathcal{B}^{'}}(\tau)=\frac{\widetilde{N}*\widetilde{\mathcal{A}}}{1+\widetilde{\mathcal{A}}*\tau}+\widetilde{\mathcal{D}}
\end{equation}

\subsubsection{Two-Measurement-Based Prediction}

We can see from Eq (11) that the equivalent aggregate request rate $\widetilde{\mathcal{A}}$ and the overall query rate from full clients $\widetilde{\mathcal{D}}$ are both unknown. Assuming the constant query pattern of stub clients and the constant requesting caching resolvers, the number of requesting caching resolvers $N$ can be estimated by the authoritative servers. Since both the caching resolvers and the full clients are visible to the authoritative servers, the authoritative servers can no longer infer the number of requesting caching resolvers simply by counting the requestors. What can differentiate a requesting caching resolver from a full client is that the query rate of a full client is hardly impacted by the variant TTL while a TTL change does impact the outbound query rate of a requesting caching resolver. So when the authoritative servers change the TTL, those requestors with comparatively minor query rate should be identified as full clients, and the number of requesting caching resolvers can be determined by subtracting the estimated amount of full clients from the total observed requestors.

To estimate the remaining two parameters, namely $\widetilde{\mathcal{A}}$ and $\mathcal{D}$, we will need the observed inbound query rates of the target authoritative servers under at least two TTL values. Since the estimation of $N$ also relies on the measurements under two TTL values, the overall measurements required by the prediction method can be summarized as the
measurements of each requestor's query rate under two TTL values. Given the two observed inbound query rates $\widetilde{\mathcal{B}^{'}}(\tau_1)$ and $\widetilde{\mathcal{B}^{'}}(\tau_2)$ under the two TTL values $\tau_1$ and $\tau_2$ and an estimated $\widetilde{N}$, $\widetilde{\mathcal{A}}$ and $\widetilde{\mathcal{D}}$ can be obtained by solving the following binary equations

\begin{equation}
\begin{cases}
\widetilde{\mathcal{B}^{'}}(\tau_1)=\frac{\widetilde{N}*\widetilde{\mathcal{A}}}{1+\widetilde{\mathcal{A}}*\tau_1}+\widetilde{\mathcal{D}}\\
\widetilde{\mathcal{B}^{'}}(\tau_2)=\frac{\widetilde{N}*\widetilde{\mathcal{A}}}{1+\widetilde{\mathcal{A}}*\tau_2}+\widetilde{\mathcal{D}}\\
\end{cases}
\end{equation}

For a new TTL of the requested record $\tau^*$, the query load of authoritative servers can be predicted using the estimation $\widetilde{\mathcal{A}}$, $\widetilde{\mathcal{D}}$ and $\widetilde{N}$:

\begin{equation}
\widetilde{\mathcal{B}}(\tau^*)=\frac{\widetilde{N}*\widetilde{\mathcal{A}}}{1+\widetilde{\mathcal{A}}*\tau^*}+\widetilde{\mathcal{D}}
\end{equation}

\begin{figure*}[!t]
\centering
\subfigure[Exponential distribution]{
\includegraphics[width=1.5in]{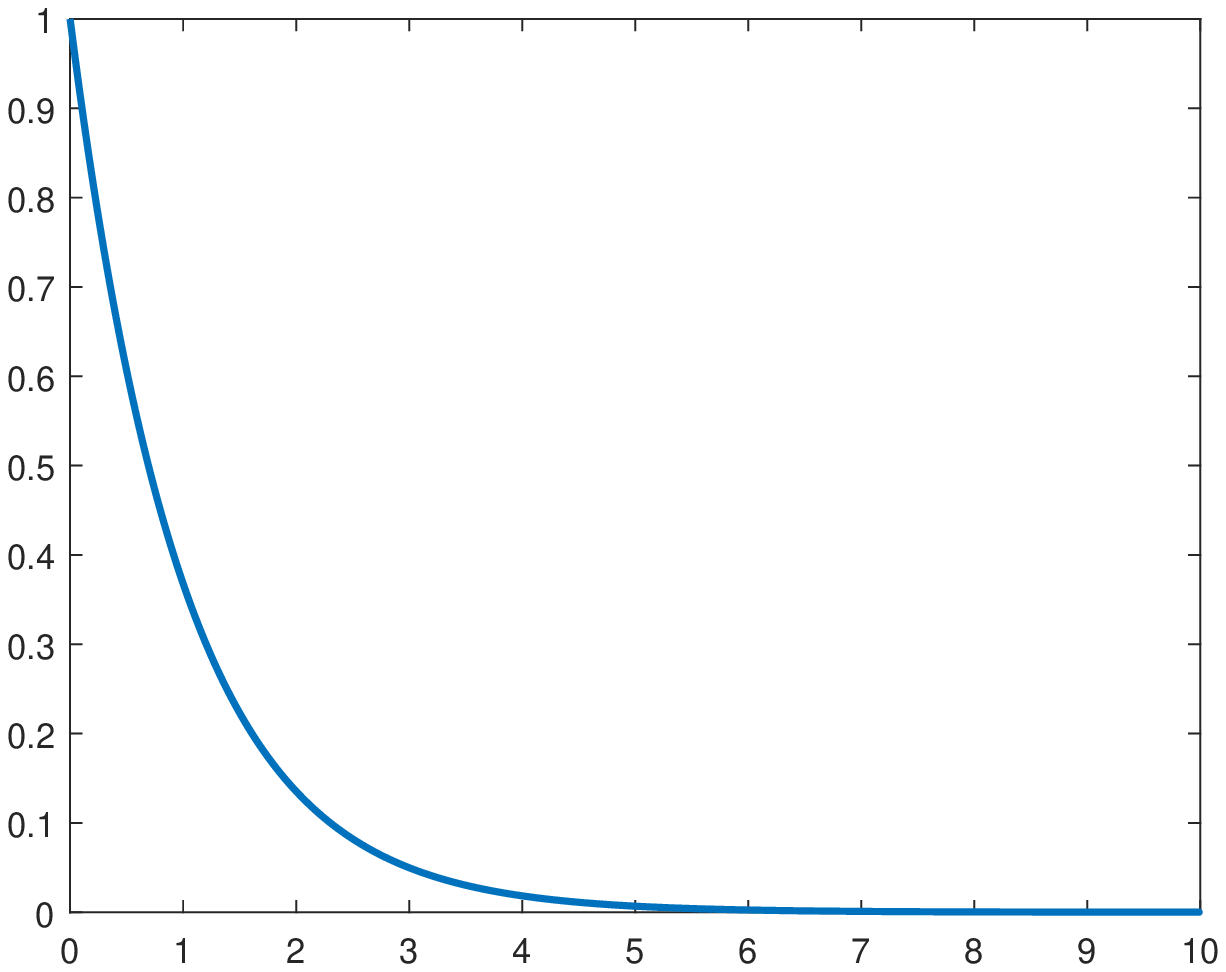}
}
\hspace{0.1in}
\subfigure[Lognormal distribution]{
\includegraphics[width=1.5in]{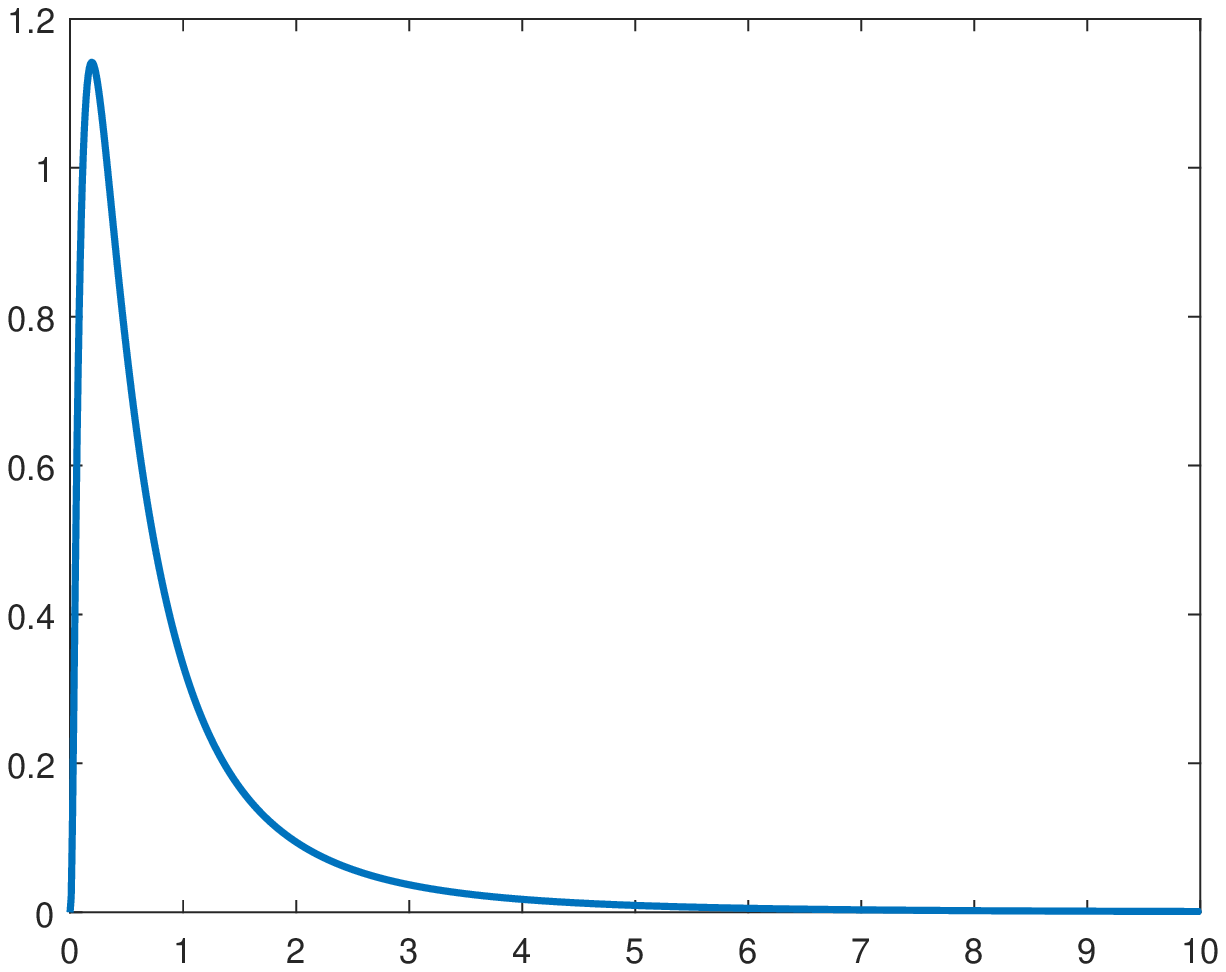}
}
\hspace{0.1in}
\subfigure[Weibull distribution]{
\includegraphics[width=1.5in]{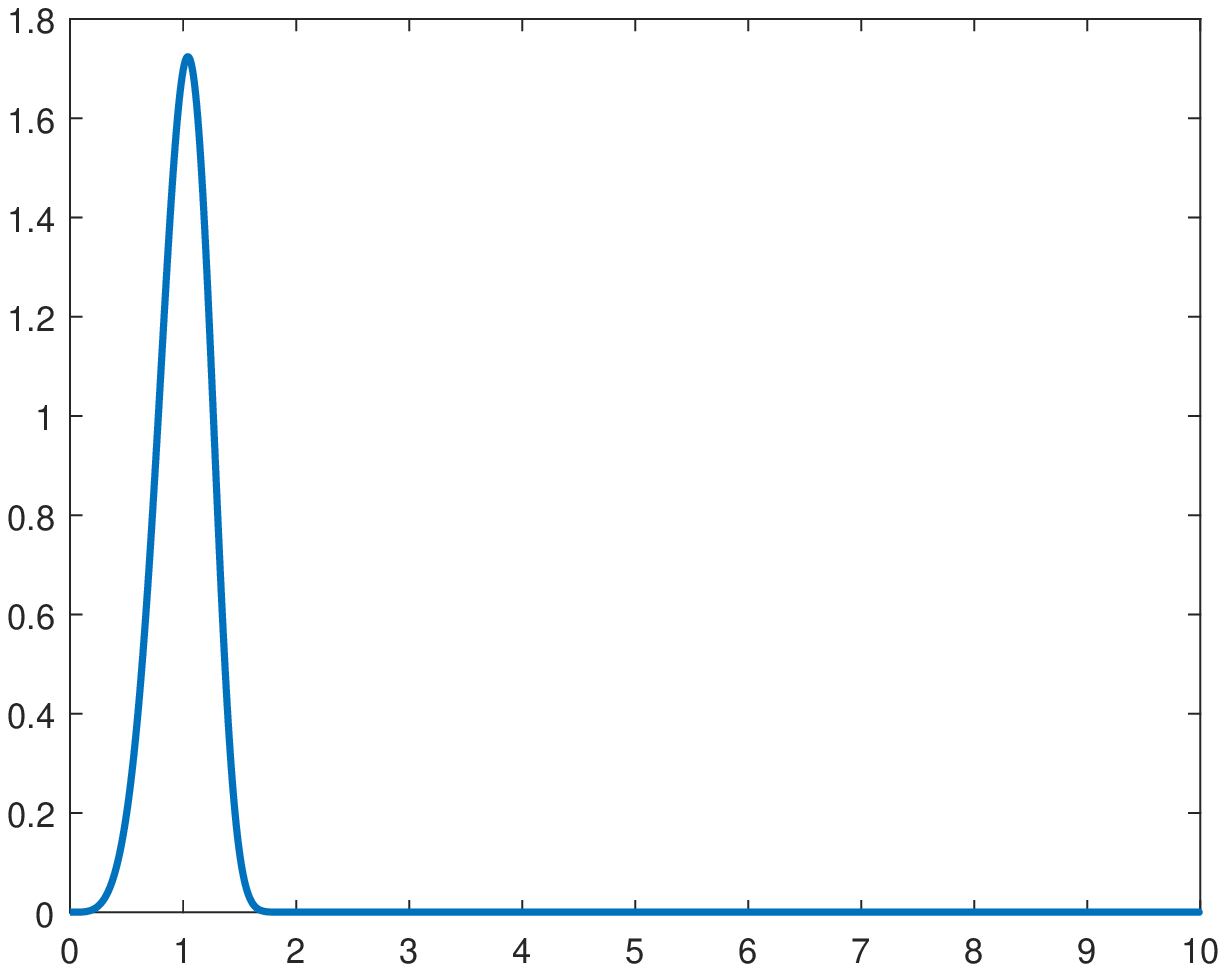}
}
\hspace{0.1in}
\subfigure[Zipf's distribution]{
\includegraphics[width=1.5in]{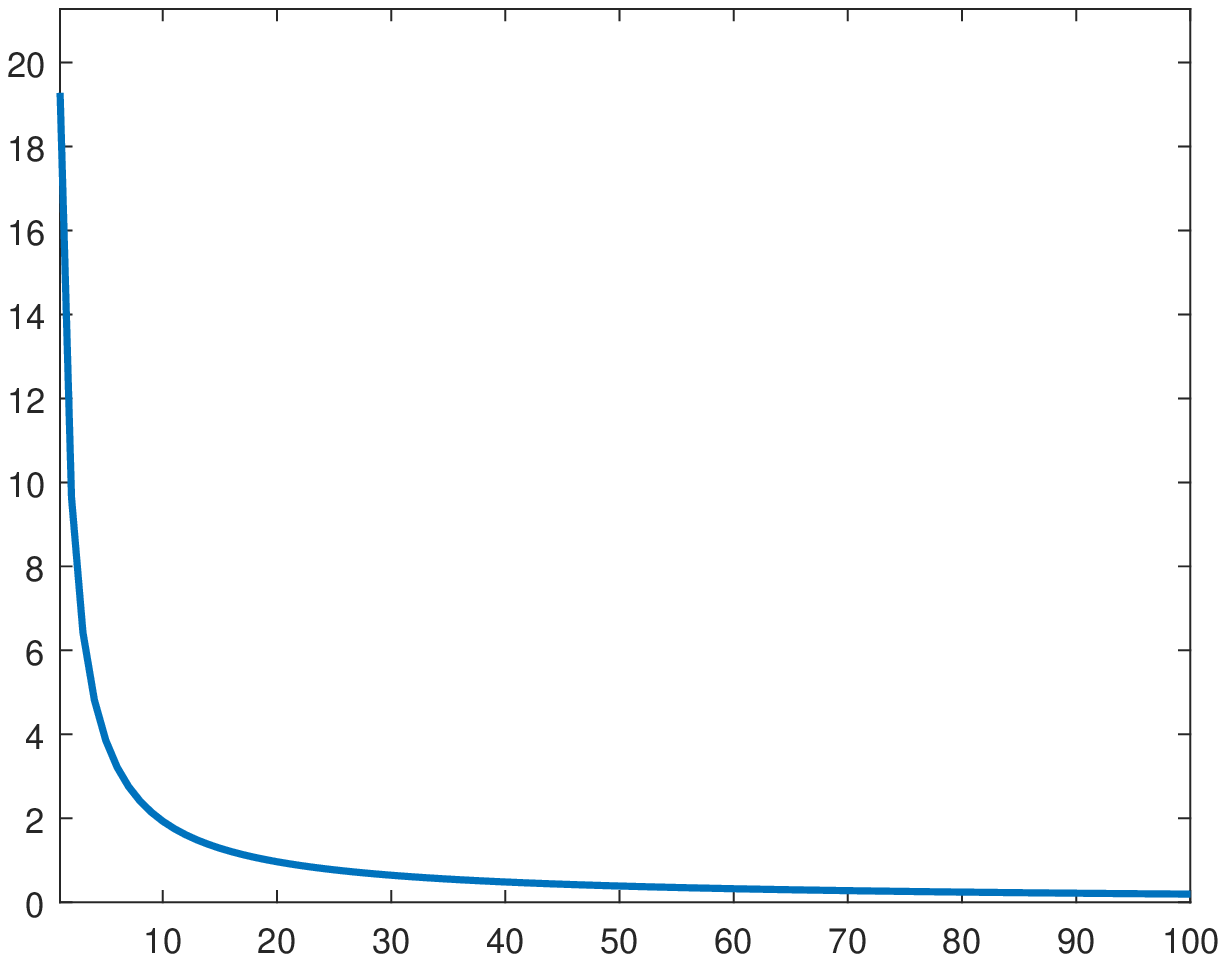}
}
\caption{PDFs of the query distributions used in validations}
\end{figure*}

\subsubsection{Three-Measurement-Based Prediction}
The accuracy of the two-measurement-based prediction is somewhat limited by the estimation $\widetilde{N}$. As mentioned above, the caching resolvers and the clients can be identified using cluster analysis with two cluster centers. The cluster of caching resolvers has a pattern of significant query rate difference between two TTL values, and the cluster of clients has a pattern of minor query rate difference. The distance between the two clusters decreases when the two TTL values become closer, and the difficulty of differentiating the two clusters adds. So the estimation $\widetilde{N}$ is prone to bigger error for close TTL values. However, if more than two measurements under different TTL values are available, the estimation $\widetilde{N}$ can be obtained by solving the three equations. So do the estimation $\widetilde{\mathcal{A}}$, $\widetilde{\mathcal{D}}$. The three equations are listed as

\begin{equation}
\begin{cases}
\widetilde{\mathcal{B}^{'}}(\tau_1)=\frac{\widetilde{N}*\widetilde{\mathcal{A}}}{1+\widetilde{\mathcal{A}}*\tau_1}+\widetilde{\mathcal{D}}\\
\widetilde{\mathcal{B}^{'}}(\tau_2)=\frac{\widetilde{N}*\widetilde{\mathcal{A}}}{1+\widetilde{\mathcal{A}}*\tau_2}+\widetilde{\mathcal{D}}\\
\widetilde{\mathcal{B}^{'}}(\tau_3)=\frac{\widetilde{N}*\widetilde{\mathcal{A}}}{1+\widetilde{\mathcal{A}}*\tau_3}+\widetilde{\mathcal{D}}\\
\end{cases}
\end{equation}

Where $\widetilde{\mathcal{B}^{'}}(\tau_i)$ is the observed inbound query rate under $\tau_i$ ($i=1, 2, 3$).

For a new TTL of the requested record $\tau^*$, the query load of authoritative servers can be predicted using Eq (13).

\begin{figure}[!t]
\centering
\subfigure[Estimation error under varying TTL of measurement and TTL of estimation]{
\includegraphics[width=1.5in]{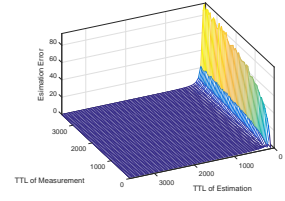}
}
\caption{Exponential distribution under stub client only model.}
\label{S-1}
\end{figure}

\begin{figure}[!t]
\centering
\subfigure[Estimation error under varying TTL of measurement and TTL of estimation]{
\includegraphics[width=1.5in]{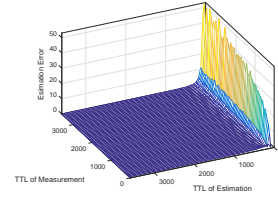}
}
\caption{Uniform distribution under stub client only model.}
\label{S-2}
\end{figure}

\section{Validations}

In this section, we use various inbound query distributions of caching resolvers to validate the proposed prediction method. For comparison, we let each of the following distribution have the mean as 1 (qps).

1) \textbf{Exponential distribution}. We use a exponential distribution with mean parameter as 1.

2) \textbf{Uniform distribution}. We use a uniform distribution with lower and upper endpoints as 0 and 2 respectively (ensuring the mean of 1).

3) \textbf{Lognormal distribution}. The probability density function (PDF) of Lognormal distribution is given by

\begin{equation}
f(x)=\frac{1}{\sqrt{2\pi}\sigma x}exp(-\frac{[ln(x)-\mu]^2}{2\sigma^2})~~~~~x>0
\end{equation}

We use a lognormal distribution with parameters $\mu$ and $\sigma$ as -0.5493 and 1.0481 respectively to ensure the mean and variance as 1 and 2 respectively.

4) \textbf{Weibull distribution}. Weibull distribution has the ability of assume the characteristics of many different types of distributions. This has made it popular among engineers. The PDF of Weibull distribution is given by

\begin{equation}
f(x)=\frac{k}{\lambda}(\frac{x}{\lambda})^{k-1}e^{-(t/\lambda)^k}~~~~~x\geq0
\end{equation}

Where $\lambda$ and $k$ are the scale and shape parameters respectively. Here we let $\lambda=1.09$ and $k=5$ ensuring the mean as 1.

%
%

5) \textbf{Zipf's distribution}. Zipf's law was first explained by G. K. Zipf \cite{zipf} who found that the frequency of any word is approximately inversely proportional to its rank in the frequency table. Zipf's law has been used to model Web links \cite{link} and network traffic \cite{traffic}. And efficient caching relies heavily on Zipf's law to replicate a small number of immensely popular files near the users \cite{cache}. Zipf's distribution is usually written as

\begin{equation}
p(k)=Ck^{-\alpha}
\end{equation}

Where the constant $\alpha \approx 1$.

\subsection{Validations under Stub Client Only Model}

\subsection{Validations under Stub Client and Full Client Model}

\subsubsection{Two-Measurement-Based Prediction:}

Let the TTL of estimation be 1800(s). For each query distribution, we first assume: 1) the number of caching resolvers and the number of full clients are both 10,000 (Requestor Distribution 1);
and then 2) 50,000 and 150,000 respectively (Requestor Distribution 2).

\begin{figure}[!t]
\centering
\subfigure[Estimation error under varying two TTLs of measurement and Requestor Distribution 1]{
\includegraphics[width=1.5in]{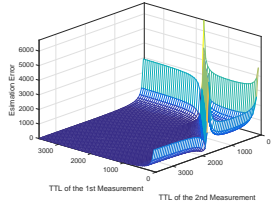}
}
\caption{Exponential distribution under stub client and full client model (two-measurement-based prediction).}
\label{S-1}
\end{figure}

\begin{figure}[!t]
\centering
\subfigure[Estimation error under varying two TTLs of measurement and Requestor Distribution 1]{
\includegraphics[width=1.5in]{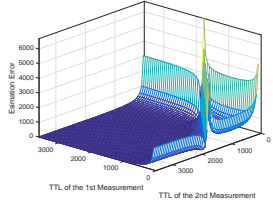}
}
\caption{Uniform distribution under stub client and full client model (two-measurement-based prediction).}
\label{S-2}
\end{figure}

\subsubsection{Three-Measurement-Based Prediction:}
Let the TTL of estimation be 1800(s) and one TTL of measurement remain constant at 1000(s). For each query distribution, we first assume: 1) the number of caching resolvers and the number of full clients are both 10,000 (Requestor Distribution 1);
and then 2) 50,000 and 150,000 respectively (Requestor Distribution 2).

\begin{figure}[!t]
\centering
\subfigure[Estimation error under varying two TTLs of measurement and Requestor Distribution 1]{
\includegraphics[width=1.5in]{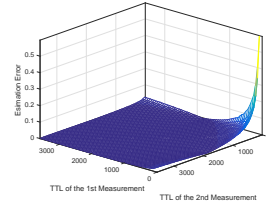}
}
\caption{Exponential distribution under stub client and full client model (three-measurement-based prediction).}
\label{S-1}
\end{figure}

\begin{figure}[!t]
\centering
\subfigure[Estimation error under varying two TTLs of measurement and Requestor Distribution 1]{
\includegraphics[width=1.5in]{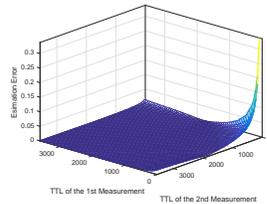}
}
\caption{Uniform distribution under stub client and full client model (three-measurement-based prediction).}
\label{S-2}
\end{figure}

\section{Global Sensitivity Analysis}

Global Sensitivity Analysis (GSA) is a term describing a set of mathematical techniques to investigate how the variation in the
output of a numerical model can be attributed to variations of its inputs. GSA can be applied for multiple purposes, including: to
apportion output uncertainty to the different sources of uncertainty of the model, e.g. unknown parameters, measurement errors
in input forcing data, etc. and thus prioritise the efforts for uncertainty reduction; to investigate the relative influence of model
parameters over the predictive accuracy and thus support model calibration, verification and simplification; to understand the
dominant controls of a system (model) and to support model-based decision-making.

In this section, we investigate how the uncertainties of the two inputs, namely the number of caching resolvers and the server load, impact the output of prediction.
We set the number of caching resolvers as 100,000, and assume a Zipf's distribution of inbound queries of caching resolvers with the mean as 1 qps.
The uncertainties of the two inputs are both configured to range between 10\% higher and lower than the true values.

%



\subsection{EET Method}

We first use the Elementary Effects Test (EET) \cite{EET} for the GSA.
The EET is a One-At-the-Time method for global Sensitivity Analysis. It computes two indices for each input: i) the mean (mi) of the EEs, which measures the total effect of an input
 over the output; ii) the standard deviation (sigma) of the EEs, which measures the degree of interactions with the other inputs. Both sensitivity indices are relative measures.



We use a One-At-the-Time sampling strategy as described in \cite{OAT}. And the sample strategy is Latin Hypercube.The base sample size is 6,000.
We use bootstrapping to derive confidence bounds. And the number of bootstrapping is 1,000.

\begin{figure}[!t]
\centering
\includegraphics[width=3.2in]{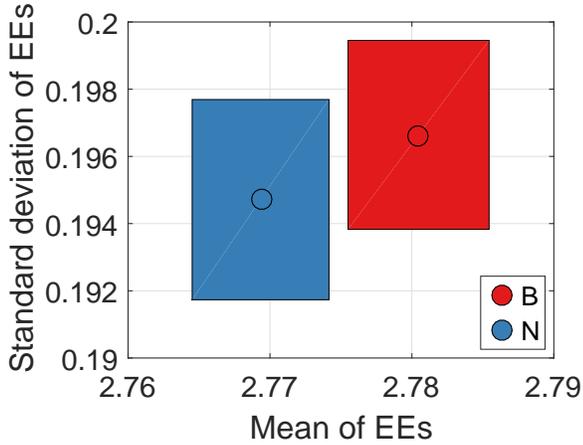}
\caption{GSA results using EET.}
\end{figure}


%
%
%
\subsection{FAST Method}

We then adopt the Fourier Amplitude Sensitivity Test (FAST) \cite{FAST} for the GSA. FAST uses the Fourier decomposition of the model output
 to approximate the variance-based first-order sensitivity indices. The base sample size is set to 3,185. The index for $N$ and $\mathcal{B}$ are 0.4902	
and 0.4935 respectively.

\subsection{VBSA Method}

We use Variance Based Sensitivity Analysis (VBSA) \cite{VBSA} for the GSA. We use two well established variance-based sensitivity indices:
the first-order sensitivity index (or 'main effects') and the total-order sensitivity index (or 'total effects'). We estimates the
main effects and the total effects indices \cite{VBSA1} by using the approximation technique described e.g. in \cite{VBSA2}. The base sample size is 6,000.
We use bootstrapping to derive confidence bounds. And the number of bootstrapping is 1,000.
\begin{figure}[!t]
\centering
\includegraphics[width=3.2in]{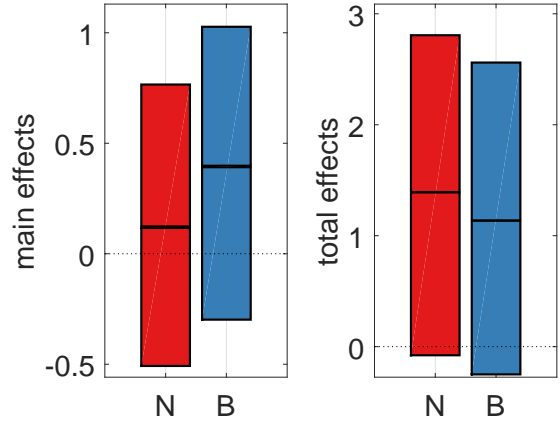}
\caption{GSA results using VBSA.}
\end{figure}

\section{Conclusion}
This paper proposed a model of DNS server load and a method of predicting DNS server load using that model. Simulations over various scenarios demonstrated that
the prediction method based on the DNS server load model has high precision and good robustness. Moreover, the prediction merely requires the limited local measurements at authoritative servers, so it
is best suited for DNS authoritative operators.


%
%




\begin{thebibliography}{1}


%

\bibitem{zipf}
G. K. Zipf, \emph{Human Behaviour and the Principle of Least Effort}. Addison-Wesley, 1949.

%
%
\bibitem{link}
M. Levene, J. Borges, and G. Loizou, Zipf's law for Web surfers. \emph{Knowledge and Information System}, 3(1), pp. 120-129, 2001.

\bibitem{traffic}
N. Sarrar, S. Uhlig, A. Feldmann, R. Sherwood, and X. Huang, Leveraging Zipf's law for traffic offloading. \emph{ACM SIGCOMM Computer Communication Review}, 42(1), pp. 16-22, 2012.

\bibitem{cache}
I. Kotera, R. Egawa, H. Takizawa, and H. Kobayashi, Modeling of cache access behavior based on Zipf's law. In \emph{Proc. of the 9th Workshop on Memory Performance: Dealing with Applications, Systems and Architecture}, pp. 9-15, 2008.

%
%

\bibitem{EET}
M. D. Morris, Factorial Sampling Plans for Preliminary Computational Experiments, \emph{Technometrics}, 33(2), pp. 161-174, 1991.

\bibitem{OAT}
F. Campolongo, A. Saltelli, and J. Cariboni, From Screening to Quantitative Sensitivity Analysis: A Unified Approach, \emph{Computer Physics
 Communications}, 182(4), pp. 978-988, 2011.

\bibitem{FAST}
R. I. Cukier, C. M. Fortuin, K. E. Shuler, A. G. Petschek, and J. H. Schaibly, Study of the Sensitivity of Coupled Reaction Systems to Uncertainties in Rate Coefficients, \emph{I Theory J Chem Phys.}, 59, pp. 3873-3878, 1973.

\bibitem{VBSA}
I. Sobol, Sensitivity Estimates for Nonlinear Mathematical Models, \emph{Mathematical Modeling \& Computational Experiment} (Engl. Transl.), 1, pp. 407-414, 1993.

\bibitem{VBSA1}
T. Homma and A. Saltelli, Importance Measures in Global Sensitivity Analysis of Nonlinear Models, \emph{Reliability Engineering \& System Safety}, 52(1), pp. 1-17, 1996.


\bibitem{VBSA2}
A. Saltelli, P. Annoni, I. Azzini, F. Campolongo, M. Ratto, and S. Tarantola, Variance Based Sensitivity Analysis of Model Output. Design and Estimator for the Total Sensitivity Index, \emph{Computer
 Physics Communications}, 181, pp. 259-270, 2010.

\bibitem{VBSA2}
A. Saltelli, P. Annoni, I. Azzini, F. Campolongo, M. Ratto, and S. Tarantola, Variance Based Sensitivity Analysis of Model Output. Design and Estimator for the Total Sensitivity Index, \emph{Computer
 Physics Communications}, 181, pp. 259-270, 2010.

\bibitem{SAFE1}
SAFE Toolbox, http://bristol.ac.uk/cabot/resources/safe-toolbox/

\bibitem{SAFE2}
F. Pianosi, F. Sarrazin, and T. Wagener, A Matlab Toolbox for Global Sensitivity Analysis, \emph{Environmental Modelling \& Software}, 70, pp. 80-85, 2015.
\end{thebibliography}
%

\end{document}